\documentstyle[12pt]{article}
\newcommand{\beq}{\begin{equation}}
\newcommand{\eeq}{\end{equation}}
\newcommand{\beqa}{\begin{eqnarray}}
\newcommand{\eeqa}{\end{eqnarray}}

\newcommand{\krig}[1]{\stackrel{\circ}{#1}}

\begin{document}

\hfill CRN 96-10

\hfill TK 96 08

\hfill hep-ph/9603nnn

\bigskip\bigskip

\begin{center}

{{\large\bf Chiral corrections to the Kroll--Ruderman theorem}}

\end{center}

\vspace{.5in}

\begin{center}
{\large V. Bernard$^a$, N. Kaiser$^b$, Ulf-G. Mei{\ss}ner$^c$}

\bigskip

\bigskip

$^a$Laboratoire de Physique Th\'eorique,
Institut de Physique \\ 3-5 rue de l'Universit\'e, F-67084 Strasbourg
Cedex, France\\
Centre de Recherches Nucl\'eaires, Physique Th\'eorique\\ 
BP 28, F-67037 Strasbourg Cedex 2, France\\ 
{\it email: bernard@crnhp4.in2p3.fr}\\
\vspace{0.3cm}
$^b$Technische Universit\"at M\"unchen, Physik Department T39\\
James-Franck-Stra\ss e, D-85747 Garching, Germany\\ 
{\it email: nkaiser@physik.tu-muenchen.de}\\
\vspace{0.3cm}
$^c$Universit\"at Bonn, Institut f{\"u}r Theoretische Kernphysik\\
Nussallee 14-16, D-53115 Bonn, Germany\\
{\it email: meissner@itkp.uni-bonn.de}\\
\end{center}

\vspace{0.5in}

\thispagestyle{empty}

\begin{abstract}
We calculate the one--loop corrections to the Kroll--Ruderman
low--energy theorems for charged pion photoproduction in the framework
of heavy baryon chiral perturbation theory. We predict the
threshold S--wave multipole $E_{0+}$ to be
$E_{0+}^{\rm thr}(\gamma p \to \pi^+ n)  = ( 28.2 \pm 0.6 )\cdot
10^{-3}/M_{\pi}$ and $E_{0+}^{\rm thr}(\gamma n \to\pi^- p) = ( -32.7 \pm
0.6 )\cdot 10^{-3}/M_{\pi}$, respectively, for a fixed
pion--nucleon coupling constant, $g_{\pi N} = 13.4$. 
A comparison to the existing data is also given.
\end{abstract}

\vfill

\today 

\newpage

\noindent {\bf 1.} Over the last years, there has been 
considerable experimental and
theoretical activitiy devoted to the subject of neutral photopion
production off protons at threshold, for reviews see e.g. \cite{dt}
\cite{bkmr}. In that reaction, the S-wave multipole $E_{0+}$ vanishes
in the chiral limit of zero pion mass and is therefore very sensitive
to the explicit chiral symmetry breaking in QCD due to the finite 
quark masses. In particular, in Refs.\cite{bkmnpb} \cite{bkmz} it was
stressed that the corresponding amplitude for the neutron is extremely
enhanced and thus should be measured. In contrast, charged pion
photoproduction at threshold is well described by the Kroll--Ruderman
term \cite{km} which is non--vanishing in the chiral limit. It
constitutes a
venerable low--energy theorem (LET) (the meaning of LETs is discussed
in \cite{cnpp})
\beqa
E_{0+}^{\rm thr} (\pi^+ n) &=& \, \, \, \, 
\frac{e \, g_{\pi N}}{4 \pi \sqrt{2} m \, (1
  + \mu)^{3/2}} = \, \, \, \, \, 27.6 \cdot 10^{-3}/M_{\pi^+} \, \, 
\, , \nonumber \\
E_{0+}^{\rm thr} (\pi^- p) &=& -\frac{e \, g_{\pi N}}{4\pi \sqrt{2} m \, (1
  + \mu)^{1/2}}= -31.7 \cdot 10^{-3}/M_{\pi^+}  \, \, \, ,
\label{e0let}
\eeqa
with $\mu = M_{\pi^+}/m$ and using $g_{\pi N}^2/4 \pi = 14.28$, $e^2 /
4 \pi = 1 /137.036$,
$m=928.27\,$MeV and $M_{\pi^+}= 139.57\,$MeV. In the limit $M_{\pi^+}
=0$, this simplifies to
\beq
E_{0+}^{\rm thr} (\pi^+ n) = -E_{0+}^{\rm thr} (\pi^- p) = 34 \cdot 
10^{-3}/M_{\pi^+} \quad .
\label{e0mpi0}
\eeq  
By comparing the numbers in Eq.(\ref{e0let}) and Eq.(\ref{e0mpi0}) one
notices that the kinematical corrections which are suppressed by
powers of the small parameter $\mu \simeq 1/7$ are quite substantial
for $E_{0+}^{\rm thr} (\pi^+ n)$. However, there are other corrections
which are related to pion loop diagrams and higher dimension operators.
These will be dealt with in a systematic fashion 
up--to--and--including order ${\cal O}(\mu^3)$ in what
follows. Before we briefly expose these calculations, let us summarize
the experimental status. Most published determinations of the 
threshold S--wave 
multipoles for charged pion photoproduction are rather old and
consistent with the Kroll--Ruderman LET,
\beqa
E_{0+}^{\rm thr} (\pi^+ n) &=& \, \, \, \,  \,
(27.9 \pm 0.5) \cdot 10^{-3}/M_{\pi^+} 
\, \cite{burg}
\, \,  ,  \, \, \, \, \, \, (28.8 \pm 0.7) \cdot 10^{-3}/M_{\pi^+} 
\, \cite{adam} \, \, \, ,
\nonumber \\
E_{0+}^{\rm thr} (\pi^- p) &=& (-31.4 \pm 1.3) \cdot 10^{-3}/M_{\pi^+} 
\, \cite{burg}
\, \,  ,  \, \, (-32.2 \pm 1.2) \cdot 10^{-3}/M_{\pi^+} \, \cite{gold} \, \, \, .
\label{data}
\eeqa
A more recent measurement of the inverse reaction $\pi^- p \to
\gamma n$ (pion radiative capture) from  TRIUMF (experiment E643) 
for energies slightly above threshold lead to the preliminary value of 
$E_{0+}^{\rm thr} (\pi^- p) = (-34.6 \pm 1.0) \cdot 10^{-3}/M_{\pi^+}$
\cite{kovash}. Here, the error is only statistical and the final
result of this experiment has not yet been reported. If it holds up,
it would amount to a rather sizeable deviation from the previously
reported numbers. 
\medskip

\noindent {\bf 2.} The tool to systematically calculate {\it all}
corrections to a given order in the pion mass is chiral perturbation
theory (CHPT). It amounts to a systematic expansion around the chiral limit
in terms of two small parameters related to the quark masses and the
external momenta. Threshold pion photoproduction is particularly
suited since these expansion parameters are given by one single number,
$M_\pi / 4 \pi F_\pi = 0.12$ (the pion energy at threshold 
is nothing but the pion mass). Here, $F_\pi =92.4\,$MeV is the pion
decay constant. We
remark that due to the presence of nucleons this small parameter does
not only appear squared as it is the case for purely mesonic processes.
Chiral corrections for charged pion photoproduction have already been 
considered in Ref.\cite{bkmnpb} within the one--loop approximation.
However, in that paper relativistic nucleon CHPT was used and thus it
could not be proven that the calculated terms of order ${\cal O}(\mu^2
\ln \mu , \mu^2)$ are not touched by higher loop corrections. This is
due to the fact that the presence of the additional mass scale related
to the nucleon mass (in the chiral limit) complicates the power
counting. This difficulty can be overcome by treating the nucleons as 
very heavy (static) sources \cite{jm}. In what follows, we will use
the systematic SU(2) approach developed in Ref.\cite{bkkm}.
The calculations are most easily done in the isospin basis,
\beq
E_{0+} (\pi^+ n)  = \sqrt{2} \, (E_{0+}^{(0)} + E_{0+}^{(-)}) \, \, ,
\, \,
E_{0+} (\pi^- p)  = \sqrt{2} \, (E_{0+}^{(0)} - E_{0+}^{(-)}) \, \, ,
\label{isobas}
\eeq
and the amplitude $E_{0+}^{(0)}$ is already known from the calculation
of the processes $\gamma N \to \pi^0 N$ since $E_{0+} (\pi^0 p/ \pi^0 n) =
\pm E_{0+}^{(0)} + E_{0+}^{(+)}$ \cite{bkmz}. In the
framework of heavy baryon CHPT, we have to consider pion loop diagrams
and local contact terms accompanied with a priori unknown coeffcients,
the so--called
low-energy constants (LECs). These we are estimating by resonance exchange
since not enough precise data exist yet to pin them all down. However,
previous calculations have already shown that this approach of
treating the LECs is fairly accurate as long as no big cancellations
appear (for details, see \cite{bkmr}). 

Consider first $E_{0+}^{(0)}$. To order $M_\pi^3$, i.e. for the 
first three terms in the chiral expansion, we find
\begin{eqnarray}
E_{0+}^{(0)} =  \frac{ C \, \mu}{2} & & 
\!\!\!\!\!\!\!\! \biggl\{ -1 +{M_\pi
\over 2m} ( 3 + \kappa_s) -{3M_\pi^2 \over 8 m^2} (5+2\kappa_s) +{M_\pi^2 \over
3\pi^2 F_\pi^2} \biggl( \ln{M_\pi\over \lambda} -{5\over6}\biggr)\nonumber  \\
& & \quad + 
{m M_\pi^2 \over 24 \pi F_\pi^2 M_K} + { (1+\kappa_\rho) m M_\pi^2  \over
 16 \pi^2 g_{\pi N} F_\pi^3 } \biggr\} 
\label{e0p0}
\end{eqnarray}
with 
\beq
C = \frac{e g_{\pi N}}{8 \pi m} = 24.01 \cdot 10^{-3}/M_{\pi^+} \, \, \,
, \label{C}
\eeq
and $\kappa_s = \kappa_p + \kappa_n = -0.12$ is the isoscalar
anomalous magnetic moment of the nucleon.
Let us briefly discuss the various contributions appearing in
Eq.(\ref{e0p0}). The first three terms come from the expansion of the
Born graphs (i.e. tree diagrams with photon absorption including the 
anomalous magnetic moment coupling followed by pion emission).
The fourth term is the pion loop contribution. Here,
$\lambda$ is the scale of dimensional regularization. We note that
the contribution $\sim M_\pi^3 \ln M_\pi$ agrees with the result of 
the relativistic calculation (as it should) \cite{bkmnpb}.
The fifth term in Eq.(\ref{e0p0}) is the
contribution from frozen kaon loops, with $M_K = 493.65 \,$MeV the
kaon  mass.  Finally, the last term stems from $\rho$--meson exchange
with $\kappa_\rho \simeq 6$ and we use some symmetry relations for the
$\rho$--meson mass and couplings \cite{bkmz}. 
These last two terms constitute the counter term contribution.
At this order, there are no other contributions to
$E_{0+}^{(0)}$.  Higher mass resonances play no role within the
accuracy of the calculation, see below.  

We turn now to the calculation of $E_{0+}^{(-)}$
to ${\cal O}(M_\pi^3)$, i.e. the first four terms in the chiral
expansion. These read
\begin{eqnarray}
E_{0+}^{(-)} = & & \!\!\!\!\!\!\!\!\!\! 
C \, \biggl\{ 1  -{M_\pi
\over m} + {9M_\pi^2 \over 8 m^2}  +{M_\pi^2 \over 8\pi^2 F_\pi^2}
\biggl({\pi^2 \over 8} - \ln{M_\pi\over \lambda}\biggr)  -{M_\pi^2
\over 16\pi^2 F_\pi^2 }\biggl({1\over 2} + \ln{M_K\over \lambda}\biggr)
\nonumber \\ 
& & \!\!\!\!\!\!\!\!\!\!  
- {M_\pi^2 \over 6} <r^2_A> + {M_\pi^2 \over 3 \sqrt{2}
m_\Delta^2} \biggl[ g_1 \biggl( 2Y(2Z-1)+{m_\Delta \over m}(1-2Y-2Z-8YZ)\biggr)
\nonumber \\ &+& g_2\biggl( (X+{1\over2})(Z-{1\over2}) -{m_\Delta \over 2m}
(1+X+Z+4XZ)\biggr) \biggr]  - {5 M_\pi^3\over 4 m^3} + {\pi \over 4}M_\pi a^+ 
\nonumber \\ &+& {M_\pi^3 \over 8 \pi^2 m F_\pi^2 } \biggl( \ln{M_\pi \over
\lambda} -{\pi^2 \over 4} - {\pi \over 2} \biggr)+ {g_{\pi N}^2 M_\pi^3 \over 
16 \pi^2 m^3 } \biggl( {\pi^2\over 8} + 1 -\ln{M_\pi \over
\lambda} \biggr) \biggr\} \, \, ,
\label{e0pm}
\end{eqnarray}
which consists of the expanded Born terms, pion loop contributions and
a variety of counter terms. These are estimated in part by $\Delta
(1232)$ excitation. The parameters related
to the $N\Delta\pi\gamma$ system ($g_1,g_2,X,Y,Z$) have been previously 
determined in chiral corrections to Compton scattering, pion--nucleon
scattering and neutral pion photoproduction \cite{bkmpi0}. 
In addition, there are
potentially large corrections related to one loop graphs with one
insertion of the dimension two operators
which come with the LECs $c_1,c_2,c_3$ and $c_4$. There are also
relativistic corrections with fixed coefficients of the type $1/2m$
and the term proportional to the isovector/isoscalar anomalous
magnetic moments $\kappa_{v,s}$ with \cite{bkmr} \cite{bkmpipin}
\beq
\kappa_v \gg 1 \, \, ,
\quad c_1,c_2,c_3, c_4 \gg \frac{1}{2m} ,\frac{g_A^2}{8m} \quad .
\eeq
However, the terms proprtional $c_{1,2,3}$ only appear in a
combination that can be expressed in terms of the small isoscalar S--wave
pion--nucleon scattering length $a^+$,
\beq
a^+ = \frac{M_\pi^2}{2\pi F_\pi^2} \biggl( -2c_1 +  c_2 + c_3 -
\frac{g_A^2}{8m} \biggr) \, \, ,
\eeq
and the other large combination $2c_4-c_3+ 1/2m$ which appears is
fully absorbed in the renormalization of the pion--nucleon vertex,
\beq
\krig{g}_A / F \to g_{\pi N} / m \quad ,
\eeq
 where $\krig{g}_A \, (F)$ denotes
the axial--vector (pion decay) constant in the chiral
limit. Furthermore, the low--energy constant $b_{11}$ related to the
Goldberger--Treiman discrepancy \cite{bkmpipin} does only enter via
the strong coupling constant renormalization.
Also, terms proportional to $\kappa_{v,s}$ appear
 only in such graphs which vanish at threshold \cite{bkmz}.
 Again, some of the terms appearing in 
Eq.(\ref{e0pm}) agree with the ones of the expanded relativistic
calculation of Ref.\cite{bkmnpb} (like e.g. the term $\sim M_\pi^2 \ln
M_\pi$). We also note that compared to that reference, we now have a
much better understanding of the $\Delta (1232)$ contribution to
certain LECs, i.e. it is much more constrained since a variety of
different processes have been calculated in the mean time.
We do not take into account isospin breaking via the difference of the
neutral and charged pion masses. Cusp effects play no role here since
the secondary thresholds ($\pi^0 p / \pi^0 n$) lie below the physical
ones.
\medskip

\noindent {\bf 3.} We are now in the position to analyse the chiral
corrections to the Kroll--Ruderman LETs. The numerical values of the
various parameters not yet  given are $X=2.75$, $Y= 0.1$, $Z=-0.2$
and $g_1 = g_2= 5$ for the $N\Delta \pi \gamma$ system \cite{bkmpi0}. We will
vary these within their bounds determined from the fit to the LEC 
$a_1 + a_2$ in neutral pion photoproduction and from the contribution 
to the $\pi N$ scattering volume $a_{33}$. For the axial 
mean square radius, we
use the dipole relation $<r_A^2> = 12/M_A^2$ with $M_A = 1.032\,$GeV.
As a variation of $M_A$, we also consider $M_A = 0.96\,$GeV and 1.15~GeV
(see e.g. \cite{choi}). In the case of  resonance saturation for the LECs,
there remains a spurious mild scale--dependence since we have to let
$\lambda$ run in the interval $M_\rho \le \lambda \le m_\Delta$.

Consider first $E_{0+}^{(0)}$. Keeping the pion--nucleon coupling
constant fixed, the Born terms are well determined. We remark that
at present there is not a generally accepted uncertainty for $g_{\pi N}$
and we thus refrain from varying it. Note, however, that
$E_{0+}^{(0)}$ essentially scales with $g_{\pi N}$ and thus a
different value than the one used here can easily be accounted for.
Letting $\lambda$ vary as described,
the pion loop contribution changes from $-0.35 \cdot
10^{-3}/M_{\pi}$ to $-0.41 \cdot 10^{-3}/M_{\pi}$. 
Further uncertainties can be estimated as follows. For
the $\rho$--contribution we change $\kappa_\rho$ from 6 to 6.6 and
in the term from the frozen kaon loops, we use $F_K = 1.21 F_\pi$
instead of $F_\pi$. Adding all these uncertainties in quadrature, 
we have ($\lambda = m$)
\beq
E_{0+}^{(0)} = ( -1.6 \pm 0.1 )\cdot 10^{-3} / M_{\pi} \quad .
\label{vale0}
\eeq
The chiral expansion is rapidly converging,
\beqa
E_{0+}^{(0)} &=& (-1.79 + 0.38 - 0.07 - 0.38 + 0.10 + 0.14) 
\cdot 10^{-3}/ M_{\pi} \,\,\, , \nonumber \\
             &=& (-1.79 + 0.38 - 0.21)  \cdot 10^{-3} /M_{\pi} \,\,\, ,
\label{expe0}
\eeqa
where the first three terms are the Born contributions of ${\cal
  O}(M_\pi^n)$ $(n=1,2,3$), while the fourth, fifth and sixth term
refer to the pion loop, the frozen K--loop and  the $\rho$-exchange 
contributions, in order. In the second line of Eq.(\ref{expe0}), we
have collected the various contributions to $E_{0+}^{(0)}$ 
of order $M_\pi$, $M_\pi^2$ and $M_\pi^3$.

Consider now $E_{0+}^{(-)}$. The pion loop contribution 
(the fourth and the last two terms in Eq.(\ref{e0pm})) is 
$(1.80,1.94,2.12) \cdot 10^{-3} / M_{\pi}$ for $\lambda = 
(M_\rho,m,m_\Delta)$ and similar for the frozen K--loop
we have $(-0.02,0.05,0.14) \cdot 10^{-3}/ M_{\pi}$. Setting
furthermore $F_K = 1.21F_\pi$, we assign a total uncertainty of $\pm 0.1  
\cdot 10^{-3} / M_{\pi}$ to this contribution.
Varying the parameters $g_2$ and $X$ under
the constraints given from $\pi^0$ photoproduction \cite{bkmpi0},
we have for the $\Delta$--contribution $(-0.57\pm 0.10) 
\cdot 10^{-3}/ M_{\pi}$. Similarly, the variation in $M_A$ leads
to $(-0.88\pm 0.17) \cdot 10^{-3}/ M_{\pi}$ from the axial radius
term. Furthermore, the term proportional to the $\pi N$ scattering
length $a^+$ induces some uncertainty. For the Karlsruhe--Helsinki
value, $a^+ = (-0.83 \pm 0.38) \cdot 10^{-2}/M_\pi$ \cite{koch},
we get a contribution  $\delta E_{0+}^{(-)} = (-0.16 \pm 0.07) \cdot
10^{-3}/ M_{\pi}$. The new value from the ETH group based on
level shifts in pionic atoms is $a^+ = (+0.25 \pm 0.18) \cdot
10^{-2}/M_\pi$  \cite{leisi}, leading to $\delta E_{0+}^{(-)} = (0.05
\pm 0.03) \cdot 10^{-3} / M_{\pi}$. Adding all these uncertainties
in quadrature, we have 
\beq
E_{0+}^{(-)} = ( 21.5 \pm 0.4 )\cdot 10^{-3} / M_{\pi} \quad . 
\label{valem}
\eeq
Higher resonance contributions are well within the given uncertainty.
It is again instructive to dissect the various terms in the chiral 
expansion,
\beq
E_{0+}^{(-)} = (24.01 - 3.57 + 1.38 - 0.29)\cdot 10^{-3}/M_{\pi} \,\,\, ,
\label{expem}
\eeq
which are the terms of ${\cal O}(M_\pi^n)$ with $n=0,1,2,3$,
respectively. Again, we find a quick convergence.

Finally, we can translate the results Eqs.(\ref{vale0},\ref{valem}) into
the physical channels,
\beqa 
E_{0+}^{\rm thr}(\gamma p \to \pi^+ n)  &=& \,\,\,\,( 28.2 \pm 0.6 )\cdot
10^{-3} / M_{\pi} \, \, , \nonumber \\
E_{0+}^{\rm thr}(\gamma n \to\pi^- p) &=& ( -32.7 \pm
0.6 )\cdot 10^{-3} / M_{\pi} \, \, ,
\label{values}
\eeqa  
for a fixed pion--nucleon coupling constant, $g_{\pi N}^2 /4 \pi =14.28$. 
These compare favorably with the existing data Eq.(\ref{data}). We
note, however, that the large (in magnitude) preliminary value from
TRIUMF for $E_{0+}^{\rm thr}(\gamma n \to\pi^- p)$ would be difficult
to understand. For comparison, in the calculation based on
relativistic CHPT we had found $E_{0+}^{\rm thr}(\gamma p \to \pi^+ n)
= 28.4 \cdot 10^{-3}/ M_{\pi}$ and  $E_{0+}^{\rm thr}(\gamma n 
\to\pi^- p) = -31.1 \cdot 10^{-3}/ M_{\pi}$ \cite{bkmnpb}.
 The differences stem
mostly from a better treatment of the $\Delta$--contribution and the
fact that in the heavy fermion approach given here, {\it all} terms of
order $M_\pi^3$ could be given. However, the statement made in \cite{bkmnpb}
that the loop corrections are fairly small in the case of charged pion
photoproduction remains valid.  

\medskip

\noindent {\bf 4.} To summarize, we have calculated the corrections to
the Kroll--Ruderman low--energy theorem up--to--and--including all
terms of order ${\cal O}(M_\pi^3)$. The chiral expansion of the
S--wave multipoles $E_{0+}^{(0)}$ and $E_{0+}^{(-)}$ shows a rapid
convergence and thus one is able to give a rather accurate prediction 
for $E_{0+}^{\rm thr}(\gamma p \to \pi^+ n)$ and $E_{0+}^{\rm
thr}(\gamma n \to \pi^- p)$, compare Eq.(\ref{values}). It would be
important to have these observables determined with high precision
for a couple of reasons. First, an accurate determination of these
multipoles gives a stringent constraint on the much discussed value
of the pion--nucleon coupling constant $g_{\pi N}$ via the
Goldberger--Miyazawa--Oehme sum rule \cite{gmo} combined with the
Panofsky ratio. Second, together with
a precise determination of the two neutral pion production amplitudes,
one would have an excellent testing ground for the investigation of 
isospin symmetry violation beyond leading order in the electromagnetic
coupling $e$. For such a test, it is mandatory to
determine the elementary neutron amplitude $\gamma n \to \pi^0 n$ (as
was stressed already in Refs.\cite{bkmnpb,bkmz}). We point out 
that the ${\cal O}(q^4)$ CHPT calculation leads us to expect that
$E_{0+}^{\rm thr}(\gamma n \to \pi^0 n) \simeq -2 \, 
E_{0+}^{\rm thr}(\gamma p \to \pi^0 p)$ 
\cite{bkmz,bkmpi0} and thus a determination using
e.g. the deuteron does appear feasible.

\end{document}